\documentstyle{article}
\newcommand{\hide}[1]{}
\begin{document}

\markboth{{\bf Are Neutrinos Majorana Particles? 
}} {{G. Rajasekaran}} 

\begin{center}
{\large{\bf ARE NEUTRINOS MAJORANA PARTICLES?}}\\
{\bf (Keynote Address)}
\vskip0.5cm
{\bf G. Rajasekaran}
\vskip0.35cm
{\it Institute of Mathematical Sciences, Madras 600113.\\
e-mail: graj@imsc.res.in}
\vskip0.35cm
\end{center}

\vspace{2cm} 

Dirac introduced the concept of antiparticles while trying
to solve the negative energy problem in the famous relativistic equation
for the electron that he had discovered in 1928. Now we know that
the existence of antiparticles is one of the most important consequences
of combining quantum mechanics with special relativity. For every particle
there exists an antiparticle. However some particles could be self-conjugate,
in the sense that particle and antiparticle could be the same. Of course such
particles have to be electrically neutral. Among the elementary particles of
the Standard Model, photon and the Z boson are self-conjugate. Both these
are bosons. The possibility of a self-conjugate fermion was first pointed
out by Majorana in 1937 and hence they are called Majorana fermions while
the other fermions (with distinct particles and antiparticles) are called Dirac Fermions. 
Among the fermions of the Standard Model, only neutrinos are
electrically neutral and hence qualify to be Majorana particles. But 
it is still an open question whether neutrinos are Majorana particles
or Dirac particles, in other words, whether neutrino is a self-conjugate particle
or not.\\  

This question can be shown to be a pure semantic one if neutrinos are massless.
In that case one can prove, by renaming the left-handed neutrino and the right-handed
antineutrino suitably, that there is no physical distinction between Dirac and
Majorana neutrinos. This is sometimes called "the confusion theorem". 
But after the discovery of neutrino mass in the last decade, we know that
such a distinction exists and physicists must determine the category to which
neutrino belongs.\\  

There are important reasons why theoreticians prefer Majorana neutrinos.
If neutrinos are Majorana particles, there exists an elegant mechanism
called sea-saw to explain why the neutrino masses, although not zero, are so tiny.
Further, if neutrinos are Majorana particles, lepton number L is not conserved
and this opens the door to generate an excess of leptons over antileptons
in the early universe which can subsequently generate an excess of baryons
over antibaryons, thus explaining how after annihilation of most of the particles with antiparticles, 
a finite but small residue of particles was left, to make up the present Universe. Hence the
fundamental importance of the question whether neutrinos are Majorana particles
is clear.\\     

Inspite of the great attractiveness of the idea that neutrinos could be
Majorana particles, it is only a theoretical idea. It has to be either
confirmed or refuted by experiment. At the present stage, the only experiment
that can answer this question is the neutrinoless double beta decay (NDBD).
In neutrinoless double beta decay,the two Majorana neutrinos that are virtually 
produced can annihilate each other
leaving only two electrons in the final state, thus violating lepton number by
two units. So establishment of the existence of this decay would be proof of
the Majorana nature of the neutrino and violation of lepton number.\\ 

Usual double beta decay (in which two neutrinos along with two electrons are emitted) 
itself is a very rare process 
because it is doubly weak as compared
to the standard beta decay and further the phase space is suppressed by the
larger number of particles in the final state. Inspite of its rarity, this decay has been
experimentally detected and now well studied. However as we saw above,
the signature for the Majorana character of the neutrino is the absence of the two neutrinos 
and this process without neutrinos in the final state has not yet been detected.
Although the neutrinoless double beta decay has reduced number of particles in the final state
and hence more phase space, it is a much rarer process than the two neutrino decay because
the decay amplitude is proportional to the tiny neutrino mass (consistent with the
confusion theorem).\\ 

Actually the decay amplitude for neutrinoless double beta decay is proportional
to a neutrino mass factor $ m_{ee}$    
which is a linear combination of the masses of the three neutrinos. This linear combination 
also involves the neutrino mixing angles and the very important CP violating
phases. Of course this has to be multiplied by the relevant nuclear matrix element
to get the decay amplitude. Thus if neutrinoless double beta decay is detected and
the rate is measured and if the nuclear matrix element is known, one can then 
extract important information on the neutrino parameters. Especially to be noted
is the fact that $m_{ee}$ contains the overall scale of the neutrino masses and
this mass-scale is not obtainable from neutrino oscillation experiments.   
However the success of such an extraction of neutrino parameters 
from NDBD depends very crucially on our
knowing the nuclear matrix elements and hence nuclear theory will play
an essential role in NDBD activity.\\ 

Can the NDBD neutrino mass factor $m_{ee}$ be zero? Unfortunately the answer
is yes. The particular linear combination of neutrino masses
that enters here may be zero (or very small) for some reason and then the rate
for NDBD may be vanishingly small. In this unfortunate case, 
NDBD experiments will not throw any light on whether neutrinos are Majorana particles.
One may have to go to processes such as a $\mu^{-}$ colliding with a nucleus
leading to a $\mu^{+}$ in the final state. The amplitude for this reaction is
proportional to a different linear combination of neutrino masses which hopefully
is not zero. But this lepton number violating reaction will be even
harder to study than NDBD.\\  

Is it possible to enhance the rates for the lepton number violating processes
by some mechanism, for instance by shining an intense laser? Such ideas are
being considered and one of them might work. But for the present let us return
to NDBD.\\ 

Inspite of the many experiments that have been mounted for searching 
for neutrinoless double beta decay, none has borne fruit sofar. However,in 2004  
Klapdor and his group 
reported observing the decay in $^{76}$Ge. This created quite
an excitement because of the importance of such a discovery. However this
result was soon controverted by many critical physicists. I was reading
Galileo's biography when this news came. 
My mind went back by 400 years to the time 
when Galileo was facing his sceptical opponents who refused to believe that
Galileo really saw the things that he claimed to have seen through his
spy glass. Of course times have changed and the situation is quite different.      
In any case there is no religious dogma concerning the nature of the neutrino!
(There is more about Galileo later in this article.)It is very important
to settle the issue by independant experiments.\\

The National Workshop on "Neutrinos in Nuclear,Particle and Astrophysics (NUPA04)"
held at IIT,Kharagpur in Feb 2004 was the first such workshop covering the impact of 
the recent discoveries of Neutrino Physics in all the
three fields of Nuclear Physics, High Energy Physics and Astrophysics. 
In that Workshop
I stressed the importance of the neutrinoless double beta decay experiment and
suggested that the Double Beta Decay Theorists Prof P K Raina and Prof P K Rath
must organize and coordinate the activity that could lead to mounting a Neutrinoless
Double Beta Decay Experiment in the INO cavern. Their magnificent enthusiasm 
led to two focussed Workshops DBD05 at IIT,Kharagpur in March 2005 and
NDBD05 at Lucknow in November 2005. More importantly, they have succeeded
in bringing two excellent experimenters Vivek Datar and RG Pillay and their groups into the
NDBD project and this augurs well for the ultimate success of
the project.We must acknowledge the important role played by the late CVK Baba
who was a pillar of strength for the project because of his wisdom and experience.
His absence is missed very much.\\

Utpal Sarkar and VKB Kota brought the experimenters
and theorists together for the NDBD07 at Ahmedabad in February 2007.
Now in October 2007, we are here for NDBD07 at Mumbai, thanks to
Vandana Nanal and her Organizing Committee.\\

\noindent {\bf Roadmap}

\vspace{2mm}

Let me close this talk with the following remarks relevant to
the roadmap ahead:\\  

1. The fundamental importance of neutrinoless double beta decay must be
stressed again and again. Repetition of this manthra is not a waste
of time since the most important unknown in all of neutrino physics is
the answer to the question whether neutrino is Majorana or Dirac. The
answer will have a bearing on High Energy Physics as well as Cosmology.
At the present time, NDBD experiment is the only way to answer it.\\

2. Hence at the India-based Neutrino Observatory (INO), NDBD must be
pursued with full vigour. Although the major focus of the INO as of now
is to construct the magnetised iron calorimeter (ICAL) to be used for
atmospheric neutrinos (Phase I) and neutrinos from muon storage rings
and beta beams (Phase II), parallel processing of NDBD must go on.
NDBD activity must soon gain sufficient strength so that the NDBD
detector will become as (if not more) important as ICAL to INO. But this requires
considerable spadework and R and D.\\

3. It is absolutely essential to scout for good experimenters and augment
the NDBD group. We must scan the whole spectrum of likely candidates in
research institutions and universities inside and outside the country
and attract them to NDBD. Also we must not restrict ourselves to
HEP and NP experimenters only. We must interact with atomic physicists,
condensed matter physicists, material scientists, chemists, engineers...
All of these can make useful contributions to mounting a viable NDBD   
experiment in the country. Students must form an important component of 
the human resource that we seek for NDBD.The task is a gigantic one, but it can be
done and it must be done.\\

4. Finally we must mention that experiments on the search for dark matter
have reached very high sensitivities and are becoming capable of detecting
it, if it really exists in the form and abundance generally expected. Hence
we must include the possibility of the Indian NDBD project leading to
an Indian Dark Matter project in the future.\\

I now add a few remarks which, although not on NDBD, will be about
neutrinos and some history.\\

\noindent {\bf A Majorana Puzzle}

\vspace{2mm}

Dirac Equation has negative energy solutions. Dirac solved the negative
energy problem filling the negative energy sea (Dirac sea) using Pauli
exclusion principle. But then he had to explain the possible vacancy
or hole in the sea. So he had to predict the antiparticle:
    Hole = Antiparticle.  
This is the famous Hole Theory of Dirac.

What happens to all this, for Majorana particles? Majorana particle
also satisfies Dirac equation and so there are negative energy solutions.
How is the negative energy problem to be solved? Answer is given at the
end of the article.\\

\noindent {\bf Mossbauer Effect for Neutrinos}

\vspace{2mm}

This is a very interesting idea by R S Raghavan (R S Raghavan,hep-ph/0601079).
Consider the following two processes:
\begin{eqnarray}
^{3}H\to\bar{\nu_{e}} + ^{3}He + e^{-}(bound) \\
\bar{\nu}_{e} + ^{3}He + e^{-} (orbital)\to ^{3}H
\end{eqnarray}

The first process is the beta decay of tritium, but the electron in the
final decay product is bound to the $^{3}He$ nucleus. Hence it is a two-body
decay with the antineutrino emitted with unique energy 18.6 keV. The
second process is just the inverse in which the same 18.6 antineutrinos
are used to cause the capture reaction which is again a two-body reaction
with the initial electron as an orbital electron in $^{3}He$. For process (1)
embed $^{3}H$ in fcc metal tritide and do the same for the initial $^{3}He$
in process(2). Thus nuclear recoil is completely avoided in both processes
just as in Mossbauer effect and by using the antinuetrinos emitted in
process (1) to initiate process(2), one is achieving recoilless resonant
capture of antineutrinos. Raghavan estimates the resonance width and gets 

\begin{equation}
\frac{\Delta E}{E} \sim 2\times 10^{-17}
\end{equation}

from which he calculates the recoilless resonant capture crosssection for process (2)
to be $\sim 5\times 10^{-32}$ cm$^{2}$.
This is 10 orders of magnitude larger than the typical capture crosssection
of antineutrinos on protons which is $\sim 10^{-42}$ cm$^{2}$.

We thus have the tools to do ultraprecise very low energy neutrino experiments.
We have a monochomatic antineutrino beam from process (1) which can be detected
with a very high crosssection. Before detection, the antineutrinos can be made
to fall through a gravitational field and thus the gravitational red-shift of
$\bar{\nu_{e}}$ can be measured. Flavour oscillations in table-top experiments
with 1gm to 1 Kg materials can be observed. This is the route to Precision
Neutrino Physics. This will revolutionize Neutrino Physics. Of course all this
is possible only if the challenges involved in the physics and technology
of the embedding mentioned above can be met.\\

\noindent {\bf Directed Monoenergetic Neutrino Beam?}

\vspace{2mm}

We envisage a directed monochromatic neutrino beam of low energy.
A possible way of realising it was considered by R S Raghavan.
There exist proposals to make high energy beams of monochromatic
neutrinos by accelerating nuclei that undergo electron capture.
(See J Sato, hep-ph/0503144, J Bernabeu, hep-ph/0505054). These
will not however have the advantage of the recoilless resonant
capture. So let us
consider the bound-state beta decay again:

\begin{equation}
^{3}H\to \bar{\nu_{e}} + ^{3}He + e^{-}(bound)
\end{equation}
and let us use a magnetic field to polarize the nuclei $^{3}H$ and $^{3}He$.
Let z be the direction of the magnetic field.
We consider angular momentum conservation.
For the initial state, we have $^{3}H$ of spin 1/2 and the 1s
atomic electron of spin 1/2. By using the external magnetic field
to polarize the nucleus and the nucleus to polarize the electron through hyperfine interaction
we can prepare the initial state such that the total spin of the
atom is in the triplet state with the z-component having value $+1$.

For the final state, the $^{3}He$ nucleus also has spin 1/2, but the
two atomic electrons are in the filled 1s shell thus contributing
zero angular momentum. As for the antineutrino, in allowed beta decay
only S-wave antineutrino participates, for the higher partial waves
are suppressed by $kR \sim 10^{-4}$, where $k$ is its momentum and R is the
nuclear radius. Hence, to balance the angular momentum, the $^{3}He$
nucleus and the antineutrino must have their $z$-component of spins
$+1/2$ making up the total z-component as $+1$, which is the initial
value.

But if the z-component of the spin of the antineutrino is $+1/2$,
its momentum also has to be in the z direction, since antineutrinos
have unique helicity. Hence the antineutrinos emitted  in this process 
in the presence of a sufficiently strong magenetic field are all
emitted along the $z$-direction. In other words, we have a monoenergetic
unidirectional (anti)neutrino beam! Now combine it with the recoilless 
emission and absorption of neutrinos (Mossbauer effect), with consequent
enormous enhancement in production and detection rates, as already
described.
This is the Ultimate Neutrino Device.

One can easily imagine any number of fantastic applications with such a device.
By having pulsed polarizing magnetic fields, even neutrino-communication
through the Earth to the antipodes is possible.

Does this make sense? See the answer at the end of the article.\\

\noindent {\bf More on Galileo}

\vspace{2mm}

I have drawn the analogy of looking for the signature of Majorana
neutrinos in NDBD detector with Galileo's looking for the signature of the new 
Astronomy in his spyglass. The analogy may prove even closer in view of
the cosmological significance of the Majorana nature of neutrinos. Because of this
and also because of the colourful picture of the Galileo episode that Arthur Koestler
paints in his book "The Watershed", I am tempted to quote an excerpt from there.
After describing the great impact of Galileo's discoveries
with his optic tube on the world at large, Koestler says:

"But to understand the reactions of the small academic world in his
own country, we must also take into account the subjective effect
of Galileo's personality. Copernicus had been a kind of
invisible man throughout his life. Nobody who met the disarming
Kepler, in the flesh or by correspondence, could seriously dislike him.
But Galileo had a rare gift of provoking enmity - not the affection
alternating with rage which Tycho aroused, but the cold, unrelenting
hostility which genius plus arrogance minus humility creates among
mediocrities.

"Without this personal background, the controversy that followed the
publication of the Sidereus Nuncius would remain incomprehensible.
For the subject of the quarrel was not the significance of the
Jupiter's satellites, but their existence, which some of Italy's
most illustrious scholars flatly denied. Galileos's main academic
rival was Magini in Bologna. In the month following the publication
of the Star Messenger, on the evenings of April 24 and 25, 1610,
a memorable party was held in a house in Bologna, where Galileo
was invited to demonstrate the Jupiter moons in his spyglass.
Not one among the numerous and illustrious guests declared himself 
convinced of their existence. Father Clavius, the leading mathematician
of Rome, equally failed to see them; Cremonini, teacher of philosophy
at Padua, refused even to look into the telescope; so did his colleague 
Libri. The latter, incidentally, died soon afterward, providing Galileo
with one more opportunity to make more enemies with the much-quoted sarcasm:
"Libri did not choose to see my celestial trifles while he was on earth;
perhaps he will do so now he has gone to heaven."

"These men may have been partially blinded by passion and prejudice, but
they were not quite as stupid as they may seem. Galileo's telescope was
the best available, but it was still a clumsy instrument without fixed
mountings, and with a visual field so small that, as somebody has said,
"the marvel is not so much that he found Jupiter's moons, but that he was
able to find Jupiter itself." The tube needed skill and experience in
handling, which none of the others possessed. Sometimes a fixed star
appeared in duplicate. Moreover, Galileo himself was unable to explain
why and how the thing worked; and the Siderius nuncius was conspicuously
silent on this essential point. Thus it was not entirely unreasonable to
suspect that the blurred dots which appeared to the strained and watering eye
pressed to the spectacle-sized lense might be optical illusions in the atmosphere, 
or somehow produced by the mysterious gadget itself. This, infact, was
asserted, in a sensational pamphlet, Refutation of the Star Messenger,
published by Magini's assistant, a young fool called Martin Horky."\\

\noindent {\bf Answer to the Majorana Puzzle}

\vspace{2mm}

Majorana spinor does not have a well-defined energy eigenvalue.
It is not the eigenfunction of the single-particle Hamiltonian.
A many-particle description (field quantization) is neccessary
for a correct understanding of the majorana particle, just as in 
the case of the spin-0 Klein-Gordon particle. (This was pointed out 
in my article on Dirac in Dirac and Feynman:Pioneers in Quantum
mechanics, Edited by Ranabir Dutt and Asim K ray, Wiley Eastern
Ltd, 1993,p 9.)

The above of course has nothing to do with the well-known
Majorana puzzle: the tragic disappearance of Ettore Majorana in
1938. On Enrico Fermi's
appeal, Benito Mussolini had mobilised the whole state
resources to search for Majorana, but to no avail.
There is some evidence that Majorana joined a monastery.
Was he disappointed that he could not make important
contributions?  Would the importance of Majorana neutrinos
to physics and cosmology (that is now recognized) have
changed his mind?\\

\noindent {\bf Answer to the Unidirectionality Question}

\vspace{2mm}

The argument that led to the unidirectionality of the (anti)neutrino is
wrong. The crucial assumption in the argument was that the (anti)neutrino
was emitted in S-wave; a pure S-wave is not possible and this is a consequence
of quantum mechanics and relativity.\\

The Dirac wavefunction for a particle with well-defined momentum p
can be written as
\begin{equation}
\psi = e ^{i\vec{p}\cdot \vec{r}}  \left( \begin{array}{c}u\\v\end{array}\right)  
\end{equation}
where $u$ and $v$ are two-component spinors and $v$ is given by
\begin{equation}
 v = \frac{\vec{\sigma}\cdot \vec{p}}{E+m}u 
\end{equation}
For $m \approx 0$, E = p and so $v$ becomes
\begin{equation}
 v = \frac{\vec{\sigma}.\vec{p}}{p}u 
\end{equation}
More generally, if energy is well-defined, but direction of $\vec{p}$
is not well-defined, the above should be rewritten as
  
\begin{equation}
\psi = \left( \begin{array}{c}F(\vec{r})\\G(\vec{r})\end{array}\right)
\end{equation}
where F(r) is a 2-component wavefunction satisfying
\begin{equation}
(\nabla^{2} + p^{2})F(\vec{r}) = 0
\end{equation}
and
\begin{equation}
G(\vec{r}) = -i \frac{\vec{\sigma}\cdot \vec{\nabla}}{p}F(\vec{r}) 
\end{equation}
Because of the
presence of  $\vec{\sigma}\cdot\vec{\nabla}$ in the wavefuncion, 
it is clear that a pure $S$-wave is not
possible. Even if we take $F(r)$ to be a pure $S$-wave (no angular dependance), $G(r)$ will
contain a $P$-wave. Thus, a pure $S$-wave is not possible for a
relativistic fermion.

For an antineutrino of mass zero, we can now impose
the helicity condition
\begin{equation}
 i\frac{\vec{\sigma}\cdot\vec{\nabla}}{p}\psi = \psi~. 
\end{equation} 
It is important to note that that this is the correct way of
writing the helicity condition which is more general than
the usual condition
\begin{equation}
\frac{\vec{\sigma}\cdot \vec{p}}{p}\psi = -\psi~,  
\end{equation}
which is valid only when the momentum direction is well-defined.
One can easily verify that the wave function satisfying the
correct helicity condition of Eq.(11) takes the form

\begin{equation}
\psi = \left(
\begin{array}{c}
H(\vec{r})\\ -H(\vec{r})
\end{array}
\right)
\end{equation}
\begin{equation} 
H(\vec{r}) = (1 + i\frac{\vec{\sigma}\cdot \vec{\nabla}}{p})K(\vec{r}) 
\end{equation}
where $H(\vec{r})$ and $K(\vec{r})$ are two-component wavefunctions and
\begin{equation}
(\nabla^{2} + p^{2})K(\vec{r}) = 0. \end{equation}      
Eq.(13) gives the form of the wavefunction $\psi$ that must be kept in mind   
while discussing angular momentum conservation. 
While considering eigenstates of angular momentum, we cannot use
eigenstates of linear momentum $\vec{p}$, since they do not commute
with each other. On the other hand, as far as eigenstates of angular momentum
are concerned, pure S-wave is impossible; even if we take
$K(r)$ to be a function of r alone without any angular dependance
it is clear that the combination that occurs in $\psi$ 
will contain both S and P waves.
With P-wave present, our argument falls to the ground.

In the above discussion, we have put the neutrino mass to
be zero and this approximation is sufficient since the mass
is very small compared to its energy 18.6 keV. If we want to
consider a purely theoretical case of the neutrino kinetic energy
comparable to or smaller than its mass, a more refined analysis
is required, but in this case, even the helicity condition
(Eq.(11) or (12)) is not valid and so there is no argument
for the (anti)neutrino emission in a single direction.

Thus a correct understanding of neutrino requires both quantum
mechanics and relativity. Further, to say that the spin of the
neutrino is pointing in the direction of its motion is not
always correct. The correct statement is Eq.(11) . These points may have
some pedagogical value since they might not have appeared in 
textbooks.

In any case, we have discussed the recoilless reactions for neutrino
physics and the aborted proposal to make a low energy unidirectional
monochromatic (anti)neutrino beam, mainly to emphasize the importance
of ingenious ideas to
take neutrino physics further. What we need are a hundred crazy
ideas. Maybe one of them will work and help us to sove the neutrino
mystery.

\end{document}